\documentclass[12pt]{article}

\usepackage{amssymb,amsmath}
\usepackage{graphicx}
\usepackage{cite}

\newcommand{\be}{\begin{equation}}
\newcommand{\ee}{\end{equation}}
\newcommand{\Dlt}{\Delta}

\newcommand{\prt}{\partial}

\newcommand{\bt}{\beta}
\newcommand{\vp}{\varphi}

\newcommand{\al}{\alpha}
\newcommand{\ra}{\rightarrow}

\newcommand{\gm}{\gamma}

\newcommand{\Gm}{\Gamma}

\numberwithin{equation}{section}

\begin{document}

\begin{center}

{\Large{\bf Extrapolation of perturbation-theory expansions by self-similar 
approximants} \\ [5mm] 
 
S. Gluzman and V.I. Yukalov$^*$} \\ [3mm]

{\it 
Bogolubov Laboratory of Theoretical Physics, \\
Joint Institute for Nuclear Research, Dubna 141980, Russia \\ [3mm]
E-mail: yukalov@theor.jinr.ru}

\end{center}

\vskip 2cm

\begin{abstract}

The problem of extrapolating asymptotic perturbation-theory expansions in
powers of a small variable to large values of the variable tending to 
infinity is investigated. The analysis is based on self-similar
approximation theory. Several types of self-similar approximants are 
considered and their use in different problems of applied mathematics is
illustrated. Self-similar approximants are shown to constitute a powerful 
tool for extrapolating asymptotic expansions of different natures. 

\end{abstract}

\vskip 1cm

{\parindent=0pt

{\bf Key words}:
\vskip 0.2cm

41A29 -- Approximation with constraints;

41A46 -- Approximation by arbitrary nonlinear expressions;

40A25 -- Approximation to limiting values;

40C15 -- Function - theoretic methods;

40H05 -- Functional analytic methods of summability.

\vskip 2cm

$^*$Corresponding author: V.I. Yukalov

{\bf E-mail}: yukalov@theor.jinr.ru}

\newpage

\section{Introduction}

There exists a very old problem constantly met in various aspects of applied
mathematics, which can be formulated as follows. Very often realistic problems 
are so complicated that they do not allow for exact solutions. It is standard
for such problems to use some kind of perturbation theory
\cite{Bog1961, Gia1972, Nay1973}. Then one gets answers in terms of expansions
in powers of a small parameter, or a small variable, say for $x \ra 0$. However,
often the problem of interest corresponds not to a small variable, but, rather 
the opposite, to large values of this variable; very often it is the infinite 
limit $x \ra \infty$ that is of the most interest \cite{Kle2006}. One could find 
this limit, provided the general formula of expansion terms would be given and 
the derived expansion would produce convergent series. None of these conditions 
is usually valid. As a rule, only a few expansion terms can be derived. 
Additionally, the resulting series are divergent, being only asymptotic 
\cite{Erd1955, Har1949}. Then the question arises: how, from the knowledge of 
several terms of an asymptotic expansion at a variable $x \ra 0$ could one find 
the limit corresponding to $x \ra \infty$?

One often extrapolates small-variable expansions by means of Pad\'{e} 
approximants \cite{Bak1996}. However, the straightforward use of these  
approximants yields
$$
P_{M/N}(x) \sim x^{M-N} \qquad (x\ra\infty) \;   ,
$$
which, depending on the relation between $M$ and $N$, can tend to:

\begin{itemize} 

\item 
infinity (when $M > N$), 

\item
zero (when $M < N$), 

\item
a constant (if $M = N$).

\end{itemize}

In that sense, the limit $x \ra \infty$ is not defined. 

When the character of the large-variable limit is known, one can invoke the 
two-point Pad\'{e} approximants \cite{Bak1996}. However the accuracy of the 
latter is not high and one confronts several difficulties: 

\begin{enumerate}

\item
First of all, when constructing these approximants, one often obtains spurious 
poles yielding unphysical singularities \cite{Bak1996}, sometimes a large 
number of poles \cite{Saf1976}. 

\item
Second, there are the cases when Pad\'{e} approximants are not able to sum 
perturbation series even for small values of an expansion parameter \cite{Sim1991}. 

\item
Third, in the majority of cases, to reach a reasonable accuracy, one needs 
to have tens of terms in perturbative expansions \cite{Bak1996}, while often 
interesting problems provide only a few terms. 

\item
Fourth, defining the two-point Pad\'{e} approximants, one always meets an 
ambiguity in distributing the coefficients for deciding which of these must
reproduce the left-side expansion and which the right-side series. This 
ambiguity worsens with the increase of the approximants' orders, making it 
difficult to compose two-point Pad\'{e} tables. For the case of a few terms, 
this ambiguity makes the two-point Pad\'{e} approximants practically 
inapplicable. For example, it has been shown \cite{Sel1989} that, for the 
same problem, one may construct different two-point Pad\'{e} approximants, 
all having correct left and right-side limits, but differing from each other 
in the intermediate region by a factor of 40, which gives 1000$\%$ uncertainty. 
This demonstrates that in the case of short series the two-point Pad\'{e} 
approximants do not allow one to get a reliable description. 

\item
Fifth, the two-point Pad\'{e} approximants cannot always be used for 
interpolating between two different expansions, but only when these two 
expansions have compatible variables \cite{Bak1996}. When these expansions
have incompatible variables, the two-point Pad\'{e} approximants cannot be 
defined in principle.

\item
Finally, interpolating between two points, one of which is finite and another
is at infinity, one is able to characterize the large-variable limit of only 
rational powers \cite{Bak1996}. 

\end{enumerate}

Another method that allows for the extrapolation of divergent series is 
optimized perturbation theory, based on the introduction of control functions
defined by an optimization condition and guaranteeing the transformation of 
divergent series into convergent series \cite{Yuk1976a, Yuk1976b, Yuk2002}. 
Since 1976, when optimized perturbation theory was introduced 
\cite{Yuk1976a, Yuk1976b}, a number of variants of different control functions 
(see discussion in (\cite{Yuk1999, Yuk2002}) have been put forward.  
Kleinert \cite{Kle1993, Kle2006} variational perturbation theory, where control 
functions are introduced through a variable transformation and variational 
optimization conditions, is particularly worth mentioning. This method provides 
good accuracy for the extrapolation of weak-coupling expansions to the 
strong-coupling limit, especially when a number of terms in the weak-coupling 
perturbation theory are available \cite{Jan1995}.        

In the present paper, we address the problem of extrapolating small-variable 
asymptotic expansions to their effective strong-coupling limits by employing
another approach, based on self-similar approximation theory 
\cite{Yuk1990a, Yuk1990b, Yuk1991, Yuk1992, Yuk1993, Yuk1994, Yuk1996}.
The main difference of this approach from optimized perturbation theory
is that we possess the approximation methods without introducing control functions, 
which makes calculations essentially simpler. Self-similar approximation theory 
can be combined with Kleinert variational perturbation theory \cite{Kle2005}.
This, however, also requires the introduction of variational control functions. 
In the present paper, however, we pay most attention to considering simpler ways 
not involving control functions. 

There exists a principal problem, when one accomplishes an extrapolation in the 
case for which the exact solution is not known and only a few terms of 
weak-coupling perturbation theory are available. This is the problem of the 
reliability of the obtained extrapolation. In such a case, it is important to 
be able to do the extrapolation by several methods, comparing their 
results. If these results yield close values, this suggests that the 
extrapolation is reliable.  

In line with this idea, we aim at employing different variants of self-similar 
approximations, applying them to the same problems and comparing the results. 
If the approximants for a problem, obtained by different methods, are close to 
each other, this would suggest that the derived values are reliable.  

We consider several variants of self-similar approximants for each problem and 
show that they really are close to each other, hence they can successfully 
extrapolate asymptotic expansions, valid at $x \ra 0$, to their effective limits 
of $x \ra \infty$. We especially concentrate on the strong-coupling limit, where
approximate methods usually are the least accurate, leading to the maximal errors.
We show that, even in this least favourable situation, with just a few perturbative 
terms available, the self-similar extrapolation methods provide reasonable accuracy.
For completeness, we also show that the self-similar methods allow
us to construct the approximants displaying good accuracy in the whole region of
the studied variable. For instance, effective equations of state can be derived,
these being in good agreement with experimental data.    

The difference of the present paper from our previous publications is in the 
following.

(i) We study several types of self-similar approximants and compare their accuracy,
which allows us to draw conclusions on the reliability of the method.

(ii) A large set of examples of different natures is analyzed, demonstrating
the generality of the method of self-similar approximants and their 
effectiveness for extrapolating different functions met in various problems
of applied mathematics.

(iii) We consider a new type of approximants resulting from a double
self-similar renormalization and show how they improve the accuracy as compared
with exact results, when these are available.

(iv) We show an effective way for calculating the large-variable critical 
exponents.    

(v) The method is shown to provide good accuracy for the whole range of the 
variable. This is demonstrated by constructing the equation of state that exactly
reproduces a phenomenological equation for quantum hard spheres.

\section{Formulation of extrapolation problem}

Suppose we are interested in the behaviour of a real function $f(x)$ of a 
real variable $x \in [0,\infty)$. Also let this function be defined by a 
complicated problem that does not allow for an explicit derivation of the
form of $f(x)$. What can only be done is use some kind of perturbation theory 
yielding asymptotic expansions representing the function
\be
\label{2.1}
f(x) \simeq f_k(x) \qquad (x \ra 0)
\ee
at small values of the variable $x \ra 0$, with $k=0,1,\ldots$ being the 
perturbation order. The perturbative series of $k$-th order can be written as 
an expansion in powers of $x$ as
\be
\label{2.2}
 f_k(x) = f_0(x) \left ( 1 + \sum_{n=1}^k a_n x^n \right ) \; ,
\ee
where $f_0(x)$ is chosen so that the series in the brackets would start 
with the term one. It is convenient to define the reduced expression
\be
\label{2.3}
 \overline f_k(x) \equiv \frac{f_k(x)}{f_0(x)} = 1 + 
\sum_{n=1}^k a_n x^n \;  ,
\ee
which will be subject to self-similar renormalization. 

Note that practically any perturbative series can be represented in 
form (2.2). For instance, if we have a Laurent-type series
$$
 f_{m+k}(x) = \sum_{n=-m}^k c_n x^n \;  ,
$$
it can be transformed to (2.2) by rewriting it as  
$$
f_{m+k}(x) = \frac{c_{-m}}{x^m} \left (
 1 +  \sum_{n=1}^{m+k} a_n x^n  \right ) \;  .
$$ 

Here we consider the series in integer powers, or those that can be 
reduced to such, since this is the most frequent type of perturbation-theory
expansions. Thus, the Puiseaux expansion \cite{Pui1850} of the type
$$
f_k(t) = \sum_{n=n_0}^k c_n t^{n/m} \;   ,
$$  
where $n_0$ is an integer and $m$ is a nonzero natural number, can be 
reduced to form (1.2) by the change of the variable $ t = x^m$. It is
possible to generalize the approach to the series of the type
$$
 f_k(x) = \sum_{n}^k c_n x^{\al_n} \qquad ( \al_n < \al_{n+1} ) \;  ,
$$
with arbitrary real powers $\alpha_n$ arranged in an ascending order.
When $\alpha_n$ pertains to an ordered group, the latter expression 
corresponds to the Hahn series \cite{Ked2001, Mac1939}.

As is known, the most difficult region for approximating is that of the 
large variable, where approximants are usually the least accurate. This 
is why our main interest here will be the large-variable behaviour of the 
function, where its asymptotic form is
\be
\label{2.4}
 f(x) \simeq Bx^\bt \qquad (x\ra\infty) \; .
\ee
The constant $B$ is called the critical amplitude and the power $\beta$
is the critical exponent.

After employing the self-similar renormalization for the reduced function
(2.3), we get a self-similar approximant $\overline f_k^*(x)$, which gives a
self-similar approximant
\be
\label{2.5}
f_k^*(x) = f_0(x) \overline f_k^*(x)
\ee
for the sought function $f(x)$. Considering for the latter the limit 
$x \ra \infty$, we find the related approximation for the critical 
amplitude and critical exponent. In many cases the exponent is known from 
other arguments. Then, we need to find only the critical amplitude.

\section{Variants of self-similar approximants}

In the cases, when one can compare the derived approximants with known 
expressions, one can easily evaluate the accuracy of the approximants. 
But how could we trust the approximants, when no exact expression for the 
sought function is available? In that case, it would be desirable to have
to hand several variants of approximants in order to compare them with 
each other. If all of them give close results, this would suggest that 
the method is reliable.   
 
Several types of approximants, based on self-similar approximation theory,  
have been derived. We shall not repeat their derivation here. This can 
be found, along with all the details, in our previous publications. We 
shall just present the corresponding expressions and explain how they 
will be used for the problem of extrapolation to infinity.

\subsection{Self-similar factor approximants}

Self-similar factor approximants have been introduced in Refs. 
\cite{Glu2003, Yuk2003}. For the reduced expansion (2.3), the $k$th order
self-similar factor approximant reads as
\be
\label{3.1}
 \overline f_k^*(x) = \prod_{i=1}^{N_k} ( 1 + A_i x)^{n_i} \;  ,
\ee
where
\begin{eqnarray}
\label{3.2}
N_k = \left\{ \begin{array}{ll}
k/2 , & ~ k = 2,4,\ldots \\
(k+1)/2 , & ~ k =3,5,\ldots
\end{array} \right. 
\end{eqnarray}
and the parameters $A_i$ and $n_i$ are defined from the accuracy-through-order
procedure, by expanding expression (3.1) in powers of $x$, comparing the 
latter expansion with the given sum (2.3), and equating the like terms in 
these expansions. When the approximation order $k=2p$ is even, the above 
procedure uniquely defines all $2p$ parameters. When the approximation order 
$k=2p+1$ is odd, the number of equations in the accuracy-through-order 
procedure is $2p$ which is by one smaller than the number of parameters. Then, 
using the scale invariance arguments \cite{Yuk2007}, one sets $A_1=1$, 
thus, uniquely defining all parameters. Another way is to find one of the 
coefficients $A_i$ from the variational optimization of the approximant 
\cite{Yuk2009b}. Both these approaches give close results, though the scaling 
procedure of setting $A_1$ to one is simpler.   

With approximant (3.1), the self-similar approximant for the sought 
function (2.5) becomes
\be
\label{3.3}
 f^*_k(x) = f_0(x) \prod_{i=1}^{N_k} ( 1 + A_i x)^{n_i} \; .
\ee
If the zero-order factor has the large-variable form
\be
\label{3.4}
 f_0(x) \simeq A x^\al \qquad (x\ra\infty) \;  ,
\ee
then approximant (3.3) behaves as 
\be
\label{3.5} 
 f_k^*(x) \simeq B_k x^\bt \qquad (x\ra\infty) \;  .
\ee
Under a given exponent $\beta$, the powers $n_i$ must satisfy the equality
\be
\label{3.6}
 \bt = \al + \sum_{i=1}^{N_k} n_i \;  ,
\ee
while the critical amplitude $B$ is approximated by
\be
\label{3.7}
 B_k = A \prod_{i=1}^{N_k} A_i^{n_i} \;  .
\ee
  
It is worth stressing that the factor $f_0(x)$ in Eq. (3.3) is explicitly defined
by the perturbative expansion (2.2), so it is known. The factor approximants (3.3)
may have singularities when some $A_i$ and $n_i$ are negative. This makes it 
possible to associate such singularities with critical points and phase transitions.
Investigation of the critical points and the related critical exponents, by means 
of the factor approximants, has been done in our previous publications 
\cite{Glu2003, Yuk2003, Yuk2004, Yuk2007, Yuk2009b}.

\subsection{Self-similar root approximants}

The derivation of the self-similar root approximants can be found in
Refs. \cite{Glu1998, Yuk1998, Yuk2002}. The self-similar renormalization
of the reduced expansion (2.3) yields
\be
\label{3.8}
 R_k(x) = \left ( \left ( \left ( \ldots ( 1 + A_1 x )^{n_1} +
A_2x^2 \right )^{n_2} + A_3x^3 \right )^{n_3} + \ldots +
A_k x^k \right )^{n_k} \;  .
\ee
The $k$th order approximant for the sought function then becomes
\be
\label{3.9}
 f_k^*(x) = f_0(x) R_k(x) \;  .
\ee
In Ref. \cite{Yuk2002}, it has been rigorously proved that the parameters 
$A_i$ and $n_i$ are uniquely defined, provided that $k$ terms of the 
large-variable expansion at $x \ra \infty$ are known, and the condition
$pn_p - p + 1 = const$ holds for $p = 1,2,\ldots, k-1$. Then expression (3.8) 
leads to
\be
\label{3.10}
 R_k(x) \simeq A_k^{n_k} x^{kn_k} \qquad (x\ra\infty ) \;  .
\ee
With the given exponent $\beta$, the power $n_k$ satisfies the relation
\be
\label{3.11}
\bt = \al + kn_k    
\ee
and the $k$th order approximation for the critical amplitude is
\be
\label{3.12}
 B_k = A A_k^{n_k} \;  .
\ee

\subsection{Iterated root approximants}

Self-similar root approximants are uniquely defined when their parameters 
are prescribed by the large-variable behaviour of the sought function. However, 
if we try to find these parameters from the small-variable expansion (2.2), 
then we meet the problem of multiple solutions \cite{Yuk2004}. To avoid this 
problem, one has to impose additional conditions on the parameters. Such a 
straightforward condition would be the requirement that all $k$ terms in root
(3.8) would contribute to the large-variable amplitude \cite{Glu2010}. For 
this, it is necessary and sufficient that the internal powers $n_j$ be 
defined as
\be
\label{3.13}
 n_j = \frac{j+1}{j} \qquad ( 1 \leq j \leq k-1 ) \;  ,
\ee
with the external power related to the exponent $\beta$ as
\be
\label{3.14}
 n_k = \frac{\gm}{k} \qquad (\gm = \bt - \al ) \;  .
\ee
Then expression (3.8) becomes the iterated root approximant
\be
\label{3.15}
 R_k(x) = \left ( \left ( \left ( \ldots ( 1 + A_1 x )^2 +
A_2x^2 \right )^{3/2} + A_3x^3 \right )^{4/3} + \ldots +
A_k x^k \right )^{\gm/k} \;  ,
\ee
where all parameters $A_j$ are uniquely defined by the accuracy-through-order 
procedure.  

In the large-variable limit, Eq. (3.15) yields
\be
\label{3.16}
 R_k \simeq \frac{B_k}{A} \; x^\gm \qquad (x\ra\infty) \;  ,
\ee
with the critical amplitude
\be
\label{3.17}
B_k = A \left ( \left ( \ldots \left ( A_1^2 + A_2 \right )^{3/2}
+ A_3 \right )^{4/3} + \ldots + A_k \right )^{\gm/k} \; .
\ee

It may happen that the iterated root approximants are well defined up to an 
order $k$, after which they do not exist because some of the parameters $A_p$ 
are negative. At the same time, the higher-order terms of perturbation-theory
expansion can be available up to an order $k+p$. How then could we use these
additional terms for constructing the higher-order approximants?

\subsection{Corrected root approximants}

Corrections to the iterated root approximants (3.15), employing the higher-order 
terms, can be constructed \cite{Glu2010} by defining the corrected root 
approximants
\be
\label{3.18}
 \widetilde R_{k/p}(x) = R_k(x) C_{k/p}(x) \;  ,
\ee
with the correction function 
\be
\label{3.19}
 C_{k/p}(x) = 1 + d_{k+1} x^{k+1} \left ( \left ( \left (
\dots ( 1 + b_1 x )^2 + b_2 x^2 \right )^{3/2} + 
b_3 x^3 \right )^{4/3} + \ldots + b_{p-1} x^{p-1} 
\right )^{-(k+1)/(p-1)} \; ,
\ee
where $p>2$ and all parameters are defined from the accuracy-through-order 
procedure, when the terms of the expansion of form (3.18) are equated with
the corresponding terms of the perturbation theory expansion. Here, the critical 
exponent is defined by the iterated root approximant (3.16), so that the 
limit $x \ra \infty$ of the correction function is finite:
\be
\label{3.20}
C_{k/p}(\infty) = 1 + d_{k+1} \left ( \left (
\dots \left ( b_1^2 + b_2 \right )^{3/2} + b_3 \right )^{4/3} + 
 \ldots + b_{p-1}  \right )^{-(k+1)/(p-1)} \;   .
\ee

The corresponding approximation for the sought function takes the form
\be
\label{3.21}
 f_{k/p}^*(x) = f_0(x) \widetilde R_{k/p}(x) \;  .
\ee
Its large-variable behaviour is
\be
\label{3.22}
 f_{k/p}^*(x) \simeq B_{k/p} x^\bt \qquad (x\ra\infty) \;  ,
\ee
with the corrected critical amplitude
\be
\label{3.23}
B_{k/p} = A B_k C_{k/p}(\infty) \;   .
\ee

\subsection{Self-similar power transforms}

It is possible to get improvement of approximants by employing power 
transforms \cite{Glu2006}. For this purpose, we define the power transform 
of the reduced expansion (2.3) as
\be
\label{3.24}
P_k(x,m) \equiv \overline f_k^m(x) \;   ,
\ee
which is expanded in powers of $x$ giving
\be
\label{3.25}
 P_k(x,m) \cong \sum_{n=0}^k b_n(m) x^n \;  .
\ee
After the self-similar renormalization of expansion (3.25), we get a 
self-similar approximant $P_k^*(x,m)$. We then accomplish the inverse
transformation 
\be
\label{3.26}
 \overline F_k(x,m) = \left [ P_k^*(x,m) \right ]^{1/m} \;  .
\ee
The powers $m_k =m_k(x)$ are defined by the variational condition
\be
\label{3.27}
 \frac{\prt\overline F_k(x,m)}{\prt m} = 0 \;  .
\ee
Finally, the corresponding approximation for the sought function is given by
\be
\label{3.28}
 f_k^*(x) = f_0(x) \overline F_k(x,m_k) \; .
\ee
When we are interested in the large-variable limit, condition (3.27) reduces 
to the differentiation of only critical amplitude.

\subsection{Double self-similar approximants}

Another way of improving the accuracy is by employing the procedure of 
self-similar renormalization twice. The fact that the accuracy does improve can 
be illustrated by those examples for which exact solutions are known. 

The double renormalization is accomplished as follows. First, renormalizing 
the reduced expansion (2.3), we construct the self-similar approximants (2.5). 
The approximants $\overline f_k^*(x)$ form the approximation sequence 
$\{\overline f_k^*(x)\}$. Introducing the expansion function $x(\vp)$ by the equation
\be
\label{3.29} 
 \overline f_1^*(x) = \vp \; , \qquad x = x(\vp) \;  ,
\ee
we define
\be
\label{3.30}
 y_k(\vp) =  \overline f_k^*( x(\vp) ) \; .
\ee
By this definition, the sequence $\{y_k(\vp)\}$ is bijective to the sequence 
$\{\overline f_k^*(x)\}$. In view of Eq. (3.29), we have
\be
\label{3.31}
  y_1(\vp) = \vp \; .
\ee

Consider the sequence $\{y_k(\vp)\}$ as the trajectory of a dynamical system
in discrete time, that is, of a cascade, with the initial condition (3.31). 
Embed this approximation cascade into an approximation flow:
\be
\label{3.32}
 \{ y_k(\vp) : \; k \in \mathbb{Z}_+ \} \subset
\{ y(t,\vp) : \; t \in \mathbb{R}_+ \} \; ,
\ee
where
$$
 \mathbb{Z}_+ \equiv \{ 0, 1, 2, \ldots \} \; , \qquad
\mathbb{R}_+ \equiv [0, \infty) \; ,
$$
so that the flow trajectory passes through all points of the cascade trajectory,
\be
\label{3.33}
  y(t,\vp) = y_k(\vp) \qquad ( t = k) \;  .
\ee
The evolution equation for the flow reads as
\be
\label{3.34}
\frac{\prt}{\prt t} \; y(t,\vp) = v(y) \;   ,
\ee
with $v(y)$ being the flow velocity.

Integrating the evolution equation (3.34) gives
\be
\label{3.35}
 \int_{y_k}^{y_k^*} \frac{dy}{v(y)} = \tau_k \;  ,
\ee
where $y_k = y_k(\vp)$ and $\tau_k$ is the minimal effective time necessary 
for reaching the approximate fixed point $y_k^*(\vp)$. The latter, 
according to definition (3.30), is a twice renormalized self-similar 
approximant
\be
\label{3.36}
 y_k^*(\vp) = \overline f_k^{**}(x(\vp) ) \;  .
\ee
Keeping in mind definition (3.30) also allows us to rewrite integral (3.35) 
as
\be
\label{3.37}
  \int_{\overline f_k^*}^{\overline f_k^{**}} \frac{d\vp}{v_k(\vp)} = 
\tau_k \;  ,
\ee
where
$$
 \overline f_k^{*} = \overline f_k^{*}(x) \; , \qquad
\overline f_k^{**} = \overline f_k^{**}(x) \;  .
$$
Assuming that we reach the quasi-fixed point in one step, we may set 
$\tau_k =1$.  

Employing in the evolution integral (3.37) the Euler discretization for 
the velocity
\be
\label{3.38}
 v_k(\vp) = y_k(\vp) - \vp = \overline f_k^{*}(x(\vp)) -
\overline f_1^{*}(x(\vp))     
\ee
and calculating this integral gives the twice renormalized approximant 
for the sought function
\be
\label{3.39}
 \overline f_k^{**}(x) = f_0(x) \overline f_k^{**}(x) \;  .
\ee
The large-variable limit of the latter 
\be
\label{3.40}
\overline f_k^{**}(x) \simeq B_k^* x^\bt \qquad 
(x \ra \infty)
\ee
defines the approximate expression for the critical amplitude $B_k^*$. 
Usually, integral (3.37) can be calculated only numerically.   

In the following sections, the above methods of extrapolation will be
illustrated by a number of examples of different nature, with the emphasis
on the large-variable limit $x \ra \infty$. Analyzing these examples, 
we shall pay most attention to the possibility of obtaining accurate
approximate expressions by taking just a few terms in the small-variable 
expansions, bearing in mind that complicated realistic problems usually 
provide us with only a small number of terms of perturbation theory.

\section{Explicitly defined functions}

In order to clearly demonstrate how the method works and to show that it
really provides good accuracy, it is illustrative to start with functions 
whose explicit form is given. This will allow us to easily evaluate the 
accuracy of approximants. The consideration of such simpler cases is 
necessary before considering the complicated problems whose exact solutions 
are not known, since only then it is possible to explicitly demonstrate 
the efficiency of the method and to evaluate what accuracy of the used 
approximants should be expected.

The variable $x$ will be varying in the range 
$[0, \infty)$.

\subsection{Function-1}

Consider a function
\be
\label{4.1}
f(x) = \frac{1}{2} \left ( \sqrt{4+x} \; - \; 1 \right ) \;   ,
\ee
which is of importance because of giving the golden ratio
$$
\frac{1}{f(1)} = 1 + f(1) = 1.618034  \;  .
$$
In its small-variable expansion 
\be
\label{4.2}
f_k(x) = \sum_{n=0}^k c_n x^n
\ee
the first five coefficients are
$$
 c_0 = \frac{1}{2} \; , \qquad c_1 = \frac{1}{8} \; , \qquad
c_2 = -\; \frac{1}{128} \; , \qquad c_3 = \frac{1}{1024} \; , 
\qquad  c_4 = -\;\frac{5}{32768} \; .
$$
Here $f_0 = c_0$.

Despite its simplicity, this function expansion is not trivial, since the 
first two coefficients are positive, after which they start alternating.

The large-variable behaviour 
\be
\label{4.3}
f(x) \simeq B x^\bt = 0.5 \sqrt{x}
\ee
shows that 
$$
 B = 0.5 \; , \qquad \bt = 0.5 \;  .
$$
Using the approximants, described above, we fix the exponent $\beta$,
concentrating on the accuracy of calculating the critical amplitude.

The method of factor approximants of Sec. 3.1 yields $B_4 = 0.440$.
Power transforms of Sec. 3.5, with the factor approximants, do not 
provide essential improvement. The optimization condition (3.27) 
results in two solutions for $m$, which yields for the amplitudes
the values $0.416$ and $0.455$. The iterated root approximants of 
Sec. 3.3 give $B_2 = 0.374, B_3 = 0.385, B_4 = 0.393$. The corrected 
iterated roots of Sec. 3.4 give $B_{2/2} = 0.422$. Power transforms, 
with iterated roots again yield two solutions for $B_2$, with the values
$0.404$ and $0.433$. All these results are close to the Pad\'{e} 
approximant $P_{2/2} = 0.433$. Essential improvement of accuracy is 
achieved by the double approximants of Sec. 3.6 on the basis of the 
iterated roots, giving $B_4^* = 0.476$.

\subsection{Function-2}

Let us take a more complicated function
\be
\label{4.4} 
 f(x) = \frac{2}{\pi} \; {\rm arccot}(-x) \exp \left ( 1
- \; \frac{1}{1+x} \right ) \;  .
\ee
In expansion (4.2), using the value ${\rm arccot}(0) = \pi/2$, we have 
$$
c_0 = 1 \; , \qquad c_1 = 1.637 \; , \qquad
c_2 = 0.137 \; , \qquad c_3 = -0.364 \; , 
\qquad  c_4 = -0.064 \;   .
$$
Again $f_0 = c_0$. Here the first three coefficients are positive, while the 
next two are negative. The limit at infinity is
\be
\label{4.5}
 f(\infty) = 2e = 5.437 \;  ,
\ee
where the equality ${\rm arccot}(-\infty) = \pi$ is used. 

The irregularity in the coefficient signs makes the extrapolation more 
difficult. The factor approximants give $f_4^*(\infty) = 9.049$. Power
transforms, with the factor approximants, improve the result yielding
the limit $5.192$. Iterated roots give $R_3(\infty) = 3.399$, 
$R_4(\infty) = 3.547$. Corrected iterated roots are close to the latter 
values: $R_{2/2}(\infty) = 3.424$. Power transforms, with iterated roots, 
give two values: $3.547$ and $4.535$. As we see, the power-transformed factor 
approximants are the most accurate.

\subsection{Function-3}

Expanding the function
\be
\label{4.6}
 f(x) = \frac{{\rm arccot}(-x)}{1+ e^{-x} } \; ,
\ee
we get the coefficients
$$
 c_0 = \frac{\pi}{4} \; , \qquad 
c_1 = \frac{1}{2}\left ( 1 + \frac{\pi}{4} \right ) \; , 
\qquad
c_2 = \frac{1}{4} \; , \qquad 
c_3 = -\; \frac{1}{6 } \left ( 1 + \frac{\pi}{16} \right )\; , 
\qquad  c_4 = -\;\frac{5}{48} \;  .
$$
Here $f_0 = c_0$. Again, the first three coefficients are positive, while the 
next two are negative. The limit at infinity is
\be
\label{4.7}
 f(\infty) = \pi \;  .
\ee

As in the previous case, the irregularity in the coefficient signs makes
extrapolation difficult. For instance, Pad\'{e} approximants fail, the best
of them giving $1.414$, which is rather far from limit (4.7). The factor
approximants give $f_4^*(\infty) = 4.759$. Power-transformed factor 
approximants are more accurate, yielding the limit $3.142$. Iterated roots
are not good, with the limit $1.698$. Power-transformed iterated roots give
two solutions: $3.742$ and $2.267$. Thus, the power-transformed factor 
approximant, with the value $3.142$, is the best.

\subsection{Debye-H\"{u}kel function}

The Debye-H\"{u}kel function
\be
\label{4.8}
  D(x) = \frac{2}{x} - 
\frac{2}{x^2} \left ( 1 - e^{-x} \right ) 
\ee
appears in the theory of strong electrolytes \cite{Lan2000}. Its expansion
gives the sign-alternating coefficients
$$
c_0 = 1 \; , \qquad c_1 = - \; \frac{1}{3} \; , 
\qquad c_2 = \frac{1}{12} \; \qquad c_3 = - \; \frac{1}{60} \; ,
$$
$$
c_4 = \frac{1}{360} \; , 
\qquad c_5 = -\; \frac{1}{2520} \; \qquad c_6 = \frac{1}{20160} \; .
$$
Here $f_0 = c_0$.

The large-variable behaviour is
\be
\label{4.9}
  D(x) \simeq \frac{2}{x} \qquad ( x \ra \infty) \; .
\ee

Factor approximants give $B_4 = 1.640$. Power-transformed factor approximants
result in $B_5 = 1.779$. Corrected factor approximants yield $B_{2/2} = 1.642$. 
Iterated roots result in $B_2 = 2.449$, $B_3 = 2.229$, $B_4 = 2.127$. For 
corrected iterated roots, we have $B_{1/2} = 1.611$, $B_{1/3} = 1.841$, 
$B_{1/4} = 1.934$, $B_{2/2} = 1.130$, $B_{2/3} = 1.712$, $B_{2/4} = 1.811$. 
Power-transformed iterated roots in the fourth order give two solutions: 
$1.993$ and $2.049$. The best two-point Pad\'{e} approximant $P_{2/2}$ 
gives the critical amplitude $1.333$, which is much worse than the self-similar
approximants of the same fourth order.

\subsection{Stirling function}

The Stirling series expansion for the function
\be
\label{4.10}
 f(x) = \frac{1}{\sqrt{2\pi}} \; e^{1/x} x^{1/x} \Gm\left ( 1 +
\frac{1}{x} \right )  
\ee
can be written as
\be
\label{4.11}
f_k(x) = \frac{1}{\sqrt{x}} \left ( 1 + 
\sum_{n=1}^k a_n x^n \right ) \;  ,
\ee
with the coefficients
$$
 a_1 = \frac{1}{12} \; , \qquad a_2 = \frac{1}{288} \; 
\qquad a_3 = -\; \frac{139}{51840} \;  ,
$$
$$
 a_4 = -\;\frac{571}{2488320} \; , \qquad 
a_5 = \frac{163879}{209018880} \; \qquad 
a_6 = \frac{5246819}{75246796800} \;  ,
$$
$$
 a_7 = -\;\frac{534703531}{902961561600} \; , \qquad  
a_8 = -\; \frac{4483131259}{86684309913600} \; .
$$
Here $f_0 = 1/ \sqrt{x}$.

The limit at infinity is
\be
\label{4.12}
f(\infty) = \frac{1}{\sqrt{2\pi}} = 0.398942 \; .
\ee

Factor approximants yield the limit $f_6^*(\infty) = 0.454$. 
Power-transformed factor approximants improve the accuracy, giving 
$f_5^*(\infty) = 0.406$. Iterated roots result in $B_2 = 0.485$, $B_3 = 0.422$, 
but the fourth-order approximant is complex. Corrected iterated roots give 
the limit $B_{2/1} = 0.408$, $B_{2/2} = 0.312$, $B_{2/3} = 0.405$. Pad\'{e}
approximants are essentially worse.

\section{Functions defined through integrals}

Many functions are defined by means of integral representations. Expansions
of such functions often result in strongly divergent series. However, 
self-similar approximants provide rather accurate extrapolation from the zero
variable to its infinite limit.

\subsection{Integral-1}

Consider the integral
\be
\label{5.1}
f(x) = ( 1 + 2x) \int_0^\infty \frac{e^{-t}}{1+x^2t^2} \; dt \; .
\ee
Its expansion in powers of $x$ contains the coefficients
$$
c_0 = 1 \; , \qquad c_1 = 2 \; , \qquad
c_2 = -2 \; , \qquad c_3 = -4 \; , \qquad  c_4 = 24 \; ,
$$
$$
c_5 = 48 \; , \qquad c_6 = -720 \; , \qquad
c_7 = -1440 \; , \qquad c_8 = 40320 \; , \qquad  c_9 = 80640 \;  .
$$
The general expressions for the latter are
$$
 c_{2n} = (-1)^n (2n)! \; , \qquad c_{2n+1} = (-1)^n 2 (2n)! \;  .
$$
The limit of Eq. (5.1) at infinity is
\be
\label{5.2}
 f(\infty) = \pi \;  .
\ee

Factor approximants yield $f_4^*(\infty) = 1.965$, $f_5^*(\infty) = 2.015$,
demonstrating good numerical convergence, e.g., giving in the ninth order
the limit $3.113$. Iterated roots lead to $R_2(\infty) = 1.754$, 
$R_3(\infty) = 2.071$, but the higher-order approximants are complex. 
Power-transformed iterated roots in fourth order give two solutions, 
$1.971$ and $2.071$.  Corrected iterated roots in the fourth order
give $2.582$ and display good numerical convergence in higher orders. 
Pad\'{e} approximants of the same order are less accurate, for instance, 
$P_{2/2} = 1.875$.

\subsection{Complimentary error function}

The complimentary error function
\be
\label{5.3}
f(x) = {\rm erfc}(-x)
\ee
is expressed through the error function as
$$
 {\rm erfc}(x) \equiv 1 - {\rm erf}(x) \;  ,
$$
the error function being
$$
 {\rm erf}(x) \equiv \frac{2}{\sqrt{\pi}} \int_0^x e^{-t^2} dt \;  .
$$
Hence, function (5.3) is defined by means of the integral
$$
 {\rm erfc}(x) \equiv 
\frac{2}{\sqrt{\pi}} \int_x^\infty e^{-t^2} dt \;   .
$$
Expanding Eq. (5.3), we get the coefficients
$$
c_0 = 1 \; , \qquad c_1 = 1.12838 \; , \qquad
c_2 = 0 \; , \qquad c_3 = -0.37613 \; , \qquad  c_4 = 0 \; .
$$
The limit at infinity is
\be
\label{5.4}
f(\infty) = 2 \;  .
\ee

All self-similar approximants give close results. Thus, factor 
approximants yield $f_4^*(\infty) = 3.772$. Iterated roots give 
$R_3(\infty) = 2.382$. Power-transformed iterated roots of fourth order 
have two solutions $2.305$ and $3.739$. Taking into account more 
expansion terms results in better accuracy. Thus, $f_5^*(\infty)=2.629$.

\subsection{Integral-2}

The function
\be
\label{5.5}
f(x) = \frac{{\rm erfc}(-x)}{1+e^{-x}}
\ee
is defined through the integral representation for the complimentary 
error function considered in the previous subsection. The coefficients 
of the corresponding expansion are
$$
 c_0 = \frac{1}{2} \; , \qquad  c_1 = 1 \; , \qquad
c_2 = 1.62838 \; , \qquad c_3 = -0.41779 \; , \qquad  
c_4 = -0.23508 \;  .
$$
The limit at infinity is
\be
\label{5.6}
 f(\infty) = 2\;  .
\ee

Factor approximants overestimate the limit, yielding 
$f_3^*(\infty) = 5.052$, $f_5^*(\infty) = 3.286$. Power-transformed factor 
approximants, on the other hand, underestimate it, giving to fourth order 
$1.392$. Iterated root approximants lead to $B_3 = 1.371$, $B_4 = 1.893$. 
Power-transformed iterated roots to fourth order give the limit $1.684$. 
In the same order, Pad\'{e} approximants give $1.027$. Iterated root 
approximants here are the most accurate. 

The large-variable behaviour of functions (5.3) and (5.5) involves 
exponentials. Therefore the accuracy of approximations can be essentially 
improved by employing exponential self-similar approximants \cite{YukGlu1998}. 
However, here we limit ourselves by the analysis of approximants described 
in Sec. 3.

\subsection{Mittag-Leffler function}

A particular case of the Mittag-Leffler function
\be
\label{5.7}
 E(x) = e^{x^2} {\rm erfc}(x) \;  ,
\ee
which is expressed through the complimentary error function, appears in
the model of anomalous diffusion \cite{Pir2005}. The small-variable 
expansion yields the coefficients
$$
 c_0 = 1 \; , \qquad  c_1 = -\; \frac{2}{\sqrt{\pi}} \; , \qquad
c_2 = 1 \; , \qquad c_3 = -\; \frac{4}{3\sqrt{\pi}} \; , \qquad  
c_4 = \frac{1}{2} \;  .
$$
In the large-variable limit, one has
\be
\label{5.8}
 E(x) \simeq \frac{B}{x} \qquad (x \ra \infty) \;  ,
\ee
with the critical amplitude 
\be
\label{5.9}
 B = \frac{1}{\sqrt{\pi}} = 0.56419 \;  .
\ee

Factor approximants give in fourth order $B_4 = 0.511$. The same result 
holds for the corrected factor approximants $B_{2/2} = 0.511$. 
Power-transformed factors yield, in fourth order, the amplitude $0.541$. 
Iterated roots lead to $B_1 = 0.886$, $B_2 = 0.741$, $B_3 = 0.680$, 
$B_4 = 0.650$. Corrected iterated roots give in fourth order $0.403$. 
Power-transformed iterated roots yield three solutions, all being close 
to $0.641$. The accuracy improves, when more terms in the expansion are 
taken into account. For instance, the factor approximants in sixth order 
give $B_6 = 0.532$.

\section{Anharmonic and nonlinear models}

Divergent series often appear in applying perturbation theory to 
anharmonic and nonlinear models that are typical for many problems in 
physics and chemistry. In these problems, perturbation theory is usually 
done with respect to a parameter called the {\it coupling parameter} which 
characterizes the strength of interactions or anharmonicity of an external
field.

\subsection{Zero-dimensional anharmonic model}

This is one of the simplest models that, at the same time, demonstrates 
mathematical features typical of many problems in chemistry and physics.
The partition function of this model reads as
\be
\label{6.1} 
 Z(g) = \frac{1}{\sqrt{\pi}} \int_{-\infty}^\infty
\exp\left ( - x^2 - gx^4 \right ) \; dx \;  ,
\ee
where $g \in [0, \infty)$ is a dimensionless coupling parameter. Weak-coupling
perturbation theory yields the series 
\be
\label{6.2}
 Z_k(g) = \sum_{n=0}^k c_n g^n \;  ,
\ee
with the coefficients
$$
 c_n = \frac{(-1)^n}{\sqrt{\pi}\; n!} \; \Gm\left ( 2n +
\frac{1}{2} \right ) \;  .
$$
Explicitly, the first few coefficients are
$$
 c_0 = 1 \; , \qquad c_1 = -\; \frac{3}{4} \; , \qquad
c_2 = \frac{105}{32} \; ,  
$$
$$
c_3 = -\; \frac{3465}{128} \; , \qquad
c_4 = \frac{675675}{2048} \; .
$$
In the strong-coupling limit,
\be
\label{6.3}
 Z(g) = \simeq B g^{-1/4} \qquad (g\ra\infty ) \; ,
\ee
with 
\be
\label{6.4}
 B = 1.022765 \;  .
\ee

Fixing the exponent $\beta$, we calculate the critical amplitude $B_k$, 
comparing it with the known exact value from Eq. (6.4). Factor approximants
give to fourth order $B_4 = 0.838$. Corrected factor approximants, to the 
same order, yield $B_{2/2} = 1.131$. Iterated root approximants give 
$B_2 = 0.760$, but the higher-order approximants are complex. Corrected 
iterated roots result in $B_{2/2} = 0.678$. Power-transformed iterated 
roots of fourth order produce two solutions, $0.879$ and $0.971$. As we see,
the best accuracy is provided by the corrected factor approximants and 
power-transformed iterated roots.

\subsection{One-dimensional anharmonic oscillator}

The anharmonic oscillator is described by the Hamiltonian
\be
\label{6.5}
 \hat H = - \; \frac{1}{2} \; \frac{d^2}{dx^2} +
\frac{1}{2} \; x^2 + gx^4 \;  ,
\ee
in which $x \in (-\infty, +\infty)$ and $g$ is a positive anharmonicity 
parameter. Perturbation theory for the ground-state energy yields 
\cite{Hio1978} the series
\be
\label{6.6}
 E_k(g) = \sum_{n=0}^k c_n g^n \;  ,
\ee
with the coefficients
$$
c_0 = \frac{1}{2} \; , \qquad c_1 = \frac{3}{4} \; , \qquad
c_2 = -\; \frac{21}{8} \;  \qquad c_3 = \frac{333}{16} \; , \qquad
c_4 = -\; \frac{30885}{128} \; .
$$
The strong-coupling limit is
\be
\label{6.7}
 E(g) \simeq 0.667986 g^{1/3} \qquad
(g \ra \infty ) \;  .
\ee

Factor approximants give $B_3 = 0.750$, $B_5 = 0.725$, $B_7 = 0.712$.
Corrected factor approximants yield $B_{3/4} = 0.728$. The power-transformed
factor approximant of fourth order gives $0.681$. Iterated root 
approximants result in $B_2 = 0.572$, $B_3 = 0.855$, but the fourth-order
approximant is complex. Corrected iterated roots give $B_4 = 0.587$, and
power-transformed iterated roots, $0.665$. The latter value is the closest
to the exact amplitude in Eq. (6.7).

Comparing these results with those obtained by means of the Kleinert variational
perturbation theory \cite{Jan1995}, we see that the latter provides better 
accuracy. However, we would like to recall that our main aim in the present 
paper is to test the methods of self-similar approximation theory, without 
involving the introduction of variational or other control functions, and based 
on just a few initial terms of perturbation theory. Although, in our case, the 
accuracy is lower than in the Kleinert method, the calculations are much 
simpler.

\subsection{Scalar field theory}

Consider the so-called $m \phi^2$ quantum field theory on a $d$-dimensional
cubic lattice with lattice spacing $a$. The free energy of the system can be
expressed \cite{Ben1994} as the integral
\be
\label{6.8}
f(x) = x \exp \left \{ 2 \int_0^\infty e^{-t} \ln \left [
e^{-xt} I_0(xt) \right ] dt \right \} \;   ,
\ee
where $I_0(\cdot)$ is a modified Bessel function of zero order and 
$x = 1/ m a^2$. Expanding the integral in powers of the variable $x$ yields 
the series 
\be
\label{6.9}
 f_k(x) = x \left ( 1 + \sum_{n=1}^k a_n x^n \right ) \;  ,
\ee
with the coefficients
$$
a_1 = -2 \; , \qquad a_2 = 3 \; , \qquad a_3 = - \; \frac{10}{3} \; ,
\qquad a_4 = \frac{29}{12} \; ,
$$
$$
a_5 = - \; \frac{11}{10} \; ,\qquad a_6 = \frac{391}{180} \; \qquad
a_7 = - \; \frac{2389}{630} \;  .
$$
When passing to continuous space, one takes the limit $a \ra 0$, which means
that $x \ra \infty$. The sought continuous-space limit is
\be
\label{6.10}
  f(\infty) = \frac{e^\gm}{2\pi} = 0.28347 \; .
\ee

Factor approximants of fourth order give the limit $0.322$ and 
power-transformed factor approximants, $0.333$. Iterated root approximants 
yield $f^*_2(\infty) = 0.408$, $f^*_3(\infty) = 0.377$, $f^*_4(\infty) = 0.365$. 
Their accuracy can be improved by taking more terms in expansion (6.9), e.g.,  
$f^*_{12}(\infty) = 0.280$. Corrected iterated roots give $f^*_{2/2}(\infty) = 0.266$, 
and power-transformed iterated roots of fourth order lead to $0.356$ and $0.347$. 
The best Pad\'{e} approximant, up to fifth order, gives $P_{2/3} = 0.326$. 
For these low orders, the most accurate is the corrected root approximant 
$f^*_{2/2}(\infty) = 0.266$.

\subsection{Nonlinear Schr\"{o}dinger equation}

The nonlinear Schr\"{o}dinger equation serves as a basic tool for modelling 
several different problems, such as those of waves on the surface of a deep 
fluid \cite{Zak1968}, electromagnetic waves in fibre optics \cite{Kar2011}, 
and Bose-Einstein condensates \cite{Pet2008, Yuk2009a, Yuk2011}. For the last 
case, it is often called the Gross-Pitaevskii equation, although Bogolubov was
the first to write down this equation for Bose systems in his famous Lectures 
on Quantum Statistics published in 1949 \cite{Bog1949} and wrote on it many 
times since (see, e.g., Refs. \cite{Bog1967, Bog1970}). This equation for 
nonequilibrium superfluids was also studied in \cite{Bog1963}. The one-dimensional 
stationary nonlinear Schr\"{o}dinger equation for Bose condensed atoms in a 
harmonic trap reads 
\be
\label{6.11}
\hat H_{NLS} \psi = E\psi \;  ,
\ee
with the nonlinear Hamiltonian
\be
\label{6.12}
 \hat H_{NLS} = - \; \frac{1}{2} \; \frac{d^2}{dx^2} +
\frac{1}{2} \; x^2 + g| \psi|^2 \;  .
\ee
Here $g$ is a dimensionless coupling parameter. The energy levels can be 
represented in the form
\be
\label{6.13}
  E(g) = \left ( n + \frac{1}{2} \right ) f(g) \; ,
\ee
where $n = 0,1,2,\ldots$ is a quantum index labelling the eigenvalues. 
Employing the optimized perturbation theory for the function $f(g)$,
as in \cite{Yuk1998}, gives the expansion
\be
\label{6.14}
 f_k(g) = 1 + \sum_{n=1}^k a_n z^n  
\ee
in powers of the effective coupling 
$$
z \equiv \frac{J_n}{n+1/2} \; g \;  ,
$$
in which
$$
 J_n \equiv \frac{1}{2^n\pi n!} \int_{-\infty}^\infty
\exp \left ( -2x^2 \right ) H_n^4(x) \; dx \; ,
$$
with $H_n(\cdot)$ being a Hermite polynomial. The coefficients in 
expansion (6.14) are
$$
 a_1 = 1 \; , \qquad a_2 = - \; \frac{1}{8} \; , \qquad 
a_3 = \frac{1}{32} \; , \qquad  a_4 = - \; \frac{1}{128} \; .
$$
Then for the strong-coupling limit we have
\be
\label{6.15}
 f(g) \simeq \frac{3}{2} \; z^{2/3} \qquad ( z\ra \infty) \;  .
\ee
Hence the critical amplitude is $B = 3/2$.

Factor approximants give $B_4 = 1.496$, which is very close to $1.5$.
Corrected factor approximants, to fourth order, yield $1.451$ and 
power-transformed factor approximants, $1.477$. Iterated roots result 
in $B_2 = 1.379$, $B_3 = 1.415$, $B_4 = 1.435$. Corrected iterated roots
give $B_{2/2} = 1.492$ and power-transformed iterated roots, $1.426$. 
For the double self-similar approximant, based on iterated roots, we 
get $B_4^* = 1.498$. The latter is slightly better than the 
value $B_4 = 1.496$, given by the factor approximant, but calculating 
the doubly renormalized approximants is essentially more complicated.   
Of course, calculations, employing any of the self-similar approximants, 
are much less time consuming than the direct solution of the nonlinear 
differential equation (6.11). 

\section{Problems in many-body theory}

Perturbation theory in many-body problems is usually accomplished with 
respect to the coupling parameter characterizing the interaction 
strength. However, this coupling parameter is often rather large. 
Moreover, perturbative expansions practically always yield divergent 
series for any finite value of the coupling parameter. Another difficulty
is that the many-body problems, as a rule, are so much complicated that
they allow one to calculate only a few low-order terms of perturbation 
theory. We show here that self-similar approximants allow for an 
effective extrapolation of such short series, giving good accuracy even 
in the extreme case of infinitely strong coupling.

\subsection{Lieb-Liniger Bose gas}

Lieb and Liniger \cite{Lie1963} have considered a one-dimensional Bose 
gas with contact interactions. The ground-state energy of the gas can be 
written as an expansion with respect to the coupling parameter as
\be
\label{7.1}
 E(g) \simeq g - \; \frac{4}{3\pi} \; g^{3/2} +
\frac{1.29}{2\pi^2} \; g^2 - 0.017201 g^{5/2} \;  .
\ee 
In the strong-coupling limit, we have the Tonks-Girardeau expression
\be
\label{7.2}
 E(\infty) = \frac{\pi^2}{3} = 3.289868 \; .
\ee
By the change of the variables
\be
\label{7.3}
e(x)  \equiv E\left ( x^2 \right ) \; , \qquad
g \equiv x^2 \;
\ee
expansion (7.1) reduces to the form
\be
\label{7.4}
  e(x) \simeq x^2 \left ( 1 + a_1 x + a_2 x^2
+ a_3 x^3 \right ) \; ,
\ee
in which
$$
a_1 = - \; \frac{4}{3\pi}=- 0.424413 \; , \qquad
a_2 = \frac{1.29}{2\pi^2} = 0.065352 \; , \qquad
a_3=-0.017201 \;   .
$$
The fourth-order term can be set as having $a_4 = 0$.

Different self-similar approximants yield close results. The most accurate
among them correspond to iterated root approximants displaying fast 
numerical convergence: $E^*_2(\infty) = 8.713$, $E^*_3(\infty) = 4.765$,
$E^*_4(\infty) = 3.2924$. The last expression provides very good accuracy,
when compared with the exact value (7.2).

\subsection{Bose-Einstein condensation temperature}

The Bose-Einstein condensation temperature of ideal uniform Bose gas in
three-dimensional space is known to be
\be
\label{7.5}
 T_0 = \frac{2\pi\hbar^2}{mk_B} \left [
\frac{\rho}{\zeta(3/2)} \right ]^{2/3} \;  ,
\ee
where $m$ is atomic mass and $\rho$, gas density. The ideal gas is, however, 
unstable below the condensation temperature \cite{Yuk2011}. Atomic 
interactions stabilize the system and shift the transition temperature
by the amount
\be
\label{7.6}
 \Dlt T_c \equiv T_c - T_0 \;  .
\ee
This shift, at asymptotically small gas parameter
\be
\label{7.7}
 \gm \equiv \rho^{1/3} a_s \;  ,
\ee
in which $a_s$ is atomic scattering length, behaves as
\be
\label{7.8}
 \frac{\Dlt T_c}{T_0} \simeq c_1 \gm \qquad ( \gm \ra 0 ) \;  .
\ee
Monte Carlo simulations \cite{Arn2001a, Arn2001b, Kas2001, Pro2001, Nho2004} 
give 
\be
\label{7.9}
 c_1 = 1.3. \pm 0.05 \;  .
\ee

At the same time, the coefficient $c_1$ can be defined 
\cite{Kas2004a, Kas2004b, Kas2004c} as the strong-coupling limit 
\be
\label{7.10}
c_1 = \lim_{g\ra\infty} c_1(g) \equiv B 
\ee
of a function $c_1(g)$ that is available only as an expansion in an 
effective coupling parameter,
\be
\label{7.11}
 c_1(g) \simeq b_1 g + b_2 g^2 + b_3 g^3 + b_4 g^4 + b_5 g^5 \;  ,
\ee
where  
$$
 b_1=0.223286\; , \qquad b_2=-0.0661032 \; , \qquad
b_3=0.026446 \; , $$
$$
b_4=-0.0129177 \; , \qquad b_5=0.00729073 \;  .
$$
Expansion (7.11) can be represented as
\be
\label{7.12}
c_1(g) \simeq b_1 g \left ( 1 +  a_1 g + a_2 g^2 + 
a_3 g^3 + a_4 g^4 \right ) \;  ,
\ee
with the coefficients
$$
 a_n \equiv \frac{b_{n+1}}{b_1} \qquad ( n = 1,2,3,4 ) \;  .
$$

Pad\'{e} approximants do not provide good accuracy, the best of them 
gives $c_1(\infty) = 0.985$. Factor approximants, to third order, yield 
$B_3 = 1.025$. At fourth order, factor approximants give $B_4 = 1.096$, 
if one of the parameters $A_i$ is set to one, and $1.446$, if it is 
defined by the variational procedure. On average, the latter values 
give $B_4 = 1.271$. Iterated roots result in $B_2 = 1.383$ to second order 
and $B_3 = 0.854$ to third order; the fourth-order approximant is
complex. Corrected iterated roots give 
$B_{1/2} = 0.924$, $B_{1/3} = 1.289$, $B_{2/2} = 1.309$. Power-transformed 
iterated roots give two solutions, $1.227$ and $1.388$, which on average
makes $1.308$. The corrected iterated root $B_{2/2} = 1.309$ produces the 
most accurate result, practically coinciding with that found by the
Monte Carlo simulations \cite{Arn2001a, Arn2001b, Kas2001, Pro2001, Nho2004}.
Kastening \cite{Kas2004a, Kas2004b, Kas2004c}, using the Kleinert variational 
perturbation theory involving seven loops, found the value $1.27 \pm 0.11$, 
which is close to our results.

\subsection{Unitary Fermi gas}

The ground-state energy of a dilute Fermi gas can be obtained by means 
of perturbation theory \cite{Bak1999, Ket2008} with respect to the effective 
coupling parameter
\be
\label{7.13}
g \equiv | k_F a_s | \;   ,
\ee
where $k_F$ is a Fermi wave number, and $a_s$, atomic scattering length.
This perturbation theory yields the expansion
\be
\label{7.14}
 E(g) \simeq c_0 + c_1 g + c_2 g^2 + c_3 g^3 + c_4 g^4 \; ,
\ee
with the coefficients
$$
 c_0 = \frac{3}{10} \; , \qquad c_1 = - \; \frac{1}{3\pi} \; , \qquad
c_2 = 0.055661 \;  ,
$$
$$
c_3 = -0.00914 \; , \qquad c_4 = -0.018604 \; .
$$
 
The scattering length, and, respectively, the effective coupling parameter
(7.13), can be varied by means of Feshbach resonance techniques in a rather 
wide range, including $g \ra \infty$. The latter limit corresponds to the
system called a unitary Fermi gas. Numerical calculations \cite{Car2003, Ast2004}
yield
\be
\label{7.15}
 E(\infty) = 0.132 \;  .
\ee

Expansion (7.14) can be rewritten in the form
\be
\label{7.16}
 E(g) \simeq c_0 \left ( 1 + a_1 g + a_2 g^2 + a_3 g^3 + a_4 g^4 
\right ) \;  ,
\ee
in which
$$
a_n \equiv \frac{c_n}{c_0} \qquad ( n = 1,2,3,4 ) \;   .
$$
  
Factor approximants give $E^*_4(\infty) = 0.174$ and corrected factor 
approximants, $0.143$. Power-transformed factor approximants yield $0.162$.
Iterated roots give $E^*_3(\infty) = 0.169$, $E^*_4(\infty) = 0.163$. Corrected
iterated roots result in $E^*_{1/2}(\infty) = 0.103$ and power-transformed
iterated roots, in $0.163$. Doubly renormalized iterated roots improve the
limit to $0.146$. Pad\'{e} approximants are not accurate, the best of them 
giving $P_{2/2} = 0.170$.

\subsection{One-dimensional Heisenberg antiferromagnet}

The ground-state energy of an equilibrium one-dimensional Heisenberg 
antiferromagnet can be represented \cite{Hor1984} as the infinite time 
limit for the energy $E(t)$ of a nonequilibrium antiferromagnet. At small 
time $t \ra 0$, one has an expansion
\be
\label{7.17}
 E(g) \simeq -\; \frac{1}{4} \left ( 1 + \sum_{n=1}^4 a_n t^n 
\right ) \;  ,
\ee
with the coefficients
$$
 a_1 = 4 \; , \qquad a_2 = -8 \; , \qquad 
a_3 = -\; \frac{16}{3} \; , \qquad a_4 = 64 \;  .
$$
In the other limit, this ground-state energy was calculated by Hulthen 
\cite{Hul1938} exactly:
\be
\label{7.18}
E = E(\infty) = - 0.4431 \;   .
\ee
We apply the self-similar approximations to extrapolate the small-time
expansion (7.17) to the infinite time limit $t \ra \infty$ determining
$E(\infty)$.

Factor approximants yield $E^*_4(\infty) = - 0.570$, with power-transformed 
factor approximants resulting in practically the same value. Corrected 
factor approximants give $E^*_{2/2}(\infty) = - 0.211$. Corrected iterated 
roots also underestimate the limit, giving $- 0.254$. Iterated roots give
$E^*_3(\infty) = - 0.511$, $E^*_4(\infty) = - 0.482$. Power-transformed 
iterated roots yield $- 0.475$. The best Pad\'{e} approximant is
$P_{2/2} = - 0.329$. The most accurate here is the power-transformed 
iterated root approximant $E^*_4(\infty) = - 0.475$.

\subsection{Fr\"{o}hlich optical polaron}

The ground-state energy of the Fr\"{o}hlich optical polaron, in the 
weak-coupling perturbation theory \cite{Koc1982, Sel1989} reads as
\be
\label{7.19}
 E(g) \simeq - g \left ( 1 + a_1 g + a_2 g^2 \right ) \;  ,
\ee
with the coefficients
$$
 a_1 = 1.591962 \times 10^{-2} \; , \qquad
a_2 = 0.806070 \times 10^{-3} \;  .
$$
In the strong-coupling limit, the asymptotic behaviour of the 
ground-state energy has been found by Miyake \cite{Miy1975, Miy1976}
in the form
\be
\label{7.20}
 E(g) \simeq Bg^2 \qquad ( g\ra \infty) \;  ,
\ee
with the amplitude
\be
\label{7.21}
 B = -0.108513 \;  .
\ee

Since just a few terms in the perturbative expansion are available, the
Pad\'{e} approximants are not applicable at all, yielding unreasonable 
values for the amplitude, by many orders differing from Eq. (7.21).  
Self-similar approximants give more realistic values. Thus, factor 
approximants give for the amplitude $B$ the value $0.061$ and iterated 
roots, $0.049$. The doubly renormalized iterated roots improve the 
accuracy, giving the value $0.1287$ for the amplitude.

\section{Characteristics of polymer systems}

Polymers are rather complicated molecules and are highly important in many
branches of physics and chemistry. As a rule, their characteristics 
are calculated by means of perturbation theory with respect to a small 
parameter, although in reality this parameter can be quite large. Self-similar
approximants can successfully extrapolate these characteristics to arbitrary
values of the parameters, including asymptotically large values.

\subsection{Randomly branched polymers}

Many characteristics of polymers are expressed through their structure 
factors. The structure factor of three-dimensional branched polymers is 
given \cite{Lam1990, Mil1991} by the confluent hypergeometric function
\be
\label{8.1}
 S(x) = F_1 \left ( 1\; ; \frac{3}{2}\; ; \frac{3}{2}\; x 
\right ) \;  ,
\ee
in which $x$ is a dimensionless wave-vector modulus. The long-wave 
expansion 
\be
\label{8.2}
S(x) \simeq c_0 + c_1 x + c_2 x^2 + c_3 x^3 + c_4 x^4 
\ee
contains the coefficients
$$
 c_0 = 1 \; , \qquad c_1 = -1 \; , \qquad c_2 = 0.6 \; , \qquad
c_3 = -0.257143 \; , \qquad c_4 = 0.085714 \; .
$$
In the short-wave limit, one has
\be
\label{8.3}
 S(x) \simeq \frac{B}{x} \qquad ( x \ra \infty ) \;  ,
\ee
with the amplitude
\be
\label{8.4}
 B = \frac{1}{3} \;  .
\ee

The reconstruction of the short-wave amplitude by Pad\'{e} approximants
leads to senseless negative values. Factor approximants give $B_4= 0.097$, 
and the power-transformed factors yield two solutions, $0.179$ and $0.329$.
Iterated roots, at low orders, overestimate the amplitude, giving
$B_2 = 0.745$, $B_3 = 0.642$, and $B_4 = 0.590$. The same happens for the 
power-transformed roots yielding the values close to $0.6$. However, the higher 
orders of the iterated roots converge to value (8.4). For instance, the 
seventh-order iterated root approximant gives a very good accuracy, 
with $B_7 = 0.330$.

\subsection{Fluctuating fluid string}

There exists an important class of systems, called fluid membranes 
\cite{Sei1997}, which finds wide applications in chemistry, biology, medicine,
and in a variety of technological applications. First, let us consider a model 
of a fluid string that is a cartoon of a one-dimensional membrane oscillating 
between two rigid walls \cite{Edw1965, Doi2001}. The free energy of the 
string coincides with the ground-state energy of a quantum particle in a 
one-dimensional rigid potential \cite{Kle1999, Kas2002}. This energy, as a 
function of a finite wall stiffness $g$, can be represented as
\be
\label{8.5}
 E(g) = \frac{\pi^2}{8g^2} \left ( 1 + \frac{g^2}{32} +
\frac{g}{4} \; \sqrt{ 1 + \frac{g^2}{64} } \right ) \;  .
\ee
The low-stiffness expansion results in
\be
\label{8.6}
 E_k(g) = \frac{\pi^2}{8g^2} \left ( 1 + \sum_{n=1}^k a_n g^n 
\right ) \;  ,
\ee
with the coefficients
$$
a_1 = \frac{1}{4} \; , \qquad a_2 = \frac{1}{32} \; , \qquad
a_3 = \frac{1}{512} \; , \qquad a_4 = 0 \; ,
$$
$$
a_5 = - \; \frac{1}{131072} \; , \qquad a_6 = 0 \; , \qquad
a_7 = \frac{1}{16777216} \;   .
$$
The case of interest corresponds to rigid walls, when the stiffness tends 
to infinity. For such rigid walls, the energy is
\be
\label{8.7}
 E(\infty) = \frac{\pi^2}{128} = 0.077106 \;  .
\ee

Pad\'{e} approximants are not applicable for this problem, giving negative 
values of the large-stiffness energy. Factor approximants give positive 
values, although overestimating the energy, e.g., $E^*_4(\infty) = 0.15$. Iterated 
roots yield $E^*_2(\infty) = 0.039$, $E^*_3(\infty) = 0.051$, and 
$E^*_4(\infty) = 0.058$. Corrected iterated roots give 
$E^*_{2/2} (\infty) = 0.169$ and power-transformed iterated roots, 
$E^*_4(\infty) = 0.065$. Taking more terms in the expansion improves the 
accuracy. Thus, iterated roots of higher orders yield $E^*_5(\infty) = 0.062$, 
$E^*_6(\infty) = 0.065$, and $E^*_7(\infty) = 0.067$. The most accurate result 
is obtained by employing the doubly renormalized iterated roots,
giving $E_2^{**}(\infty) = 0.07237$. The variational perturbation theory, to 
sixth order, gives \cite{Kas2006} the value $0.076991$.

\subsection{Fluctuating fluid membrane}

In the case of a two-dimensional membrane, its pressure can be calculated by
perturbation theory with respect to the wall stiffness \cite{Kas2006}, which 
yields
\be
\label{8.8}
 p_k(g) = \frac{\pi^2}{8g^2} \left ( 1 + 
\sum_{n=1}^k a_n g^n \right ) \;  ,
\ee
with the coefficients
$$
a_1 = \frac{1}{4} \; , \qquad a_2 = \frac{1}{32} \; , \qquad
a_3=2.176347\times 10^{-3} \; ,
$$
$$
 a_4=0.552721\times 10^{-4} \; , \qquad 
a_5=-0.721482\times 10^{-5} \; , \qquad 
a_6=-1.777848\times 10^{-6} \; .
$$
The rigid-wall limit, calculated by means of the Monte Carlo simulations 
\cite{Gom1989} is found to be
\be
\label{8.9}
 p(\infty) =0.0798 \pm 0.0003 \;  .
\ee

Pad\'{e} approximants are again not applicable, resulting in negative values
of pressure. Factor approximants of low orders overestimate the limit, e.g., 
the fourth order giving $0.312$. To higher orders, factor approximants become 
slightly better, but still overestimating the pressure. Iterated roots of low 
orders give $p^*_2(\infty)=0.039$,  $p^*_3(\infty)=0.053$, and 
$p^*_4(\infty)=0.061$ and power-transformed iterated roots in fourth order, 
$0.068$. Taking into account all available coefficients improves the results. 
For instance, in the case of the iterated roots, we have $p^*_5(\infty)=0.067$, 
$p^*_6(\infty)=0.071$. Doubly renormalized iterated roots give 
$p^{**}_3(\infty) = 0.0792$, which is the most accurate result. This is to be 
compared with the value of $0.0821$ from the variational perturbation theory
\cite{Kas2006}, which overestimates the Monte Carlo result (8.9).

\subsection{Two-dimensional polymer chain}

An important characteristic of polymer chains is their expansion factor, 
that is, the ratio of the mean-square end-to-end distance of the chain, with 
interactions between its segments, to the value of the mean-square end-to-end
distance of the chain, without such interactions. Two-dimensional polymers
are often met in chemistry and biology. For such polymers, perturbation theory 
with respect to weak interactions can be developed \cite{Mut1984, Mut1987} and, 
in a certain limiting case, can be reduced to a series in a single 
dimensionless interaction parameter $g$. For a two-dimensional polymer chain, 
perturbation theory results \cite{Mut1984} in the expansion factor 
\be
\label{8.10}
 F(g) \simeq 1 + \sum_{n=1}^4 a_n g^n \;  ,
\ee
with the coefficients
$$
 a_1 = \frac{1}{2} \; , \qquad a_2 = -0.12154525 \; , \qquad
a_3=0.02663136 \; , \qquad a_4=-0.13223603 \;  .
$$
In the strong-interaction limit \cite{Li1995}, one has
\be
\label{8.11}
F(g) \simeq Bg^\bt \qquad (g \ra \infty ) \;   ,
\ee
with the critical exponent
\be
\label{8.12}
  \bt = 1 \;  .
\ee
One also considers the critical index
\be
\label{8.13}
  \nu \equiv \frac{1}{2} \left ( 1 + \frac{\bt}{2} \right ) \; ,
\ee
which here is $\nu = 0.75$.

Calculating the critical amplitude, we have the following. Factor 
approximants are complex, but the power-transformed factor approximant
at fourth order gives $0.31$. Iterated roots yield $B_2 = 0.08$, with
the higher orders being complex. The corrected iterated roots yield
$B_{2/2} = 0.09$. The exact value of the amplitude $B$ is not known, 
because of which we cannot evaluate the accuracy of the approximants.
But, as we see, all approximants give the values of order $0.1$.

\subsection{Three-dimensional polymer coil}

In the case of a three-dimensional polymer coil, perturbation theory
\cite{Mut1984} for the expansion factor leads to series (8.10), however with
the coefficients
$$
a_1 = \frac{4}{3} \; , \qquad a_2 =-2.075385396 \; , \qquad
a_3 = 6.296879676 \; ,
$$
$$
a_4 = -25.05725072 \; , \qquad a_5=116.134785 \; , \qquad
a_6 = - 594.71663 \; .
$$
The strong-coupling limit \cite{Mut1987} is
\be
\label{8.14}
 F(g) \simeq 1.531 g^{0.3544} \qquad ( g \ra \infty) \;  ,
\ee
which yields for the critical index (8.13) $\nu = 0.5866$. Numerical fitting
\cite{Mut1987} for the whole range of interactions results in the formula
\be
\label{8.15}
 F(g) = \left ( 1 + 7.524 g + 11.06 g^2 \right )^{0.1772} \;  .
\ee

Employing four terms in the weak-coupling expansion gives for the factor 
approximants the amplitude $B_4 = 1.548$ and for power-transformed factor 
approximants, $1.535$. Iterated roots yield 
$B_2 = 1.543$, $B_3 = 1.549$, $B_4 = 1.538$. Corrected iterated roots result
in $B_{2/2} = 1.544$ and power-transformed iterated roots, in $B_4 = 1.535$.
Doubly renormalized iterated roots give $1.530$. Higher-order approximants
improve the results, but already at fourth order all these approximants are
close to the numerical value $B = 1.531$. The accuracy of Pad\'{e} 
approximants is several orders worse \cite{Glu2003}.

\section{Calculation of critical exponents}

In the previous sections, we have concentrated on the calculation of 
critical amplitudes, with known critical exponents, by extrapolating 
the small-variable perturbative expansions to the large-variable limit, 
employing the techniques of self-similar approximants. Now we show how the
critical exponents can also be found by using these techniques.

\subsection{Scheme of general approach}

When a function, for asymptotically large variable, behaves as
\be
\label{9.1}
 f(x) \simeq B x^\bt \qquad ( x \ra \infty ) \;  ,
\ee
then the critical exponent can be represented by the limit
\be
\label{9.2}
 \bt = \lim_{x\ra\infty} x \; \frac{d}{dx} \; \ln f(x) \;  .
\ee

Assuming that the small-variable expansion for the function is given by the
sum $f_k(x)$, as in Eq. (2.2), we have the corresponding small-variable 
expression for the critical exponent
\be
\label{9.3}
 \bt_k(x) = x \; \frac{d}{dx} \; \ln f_k(x) \;  ,
\ee
which can be expanded in powers of $x$, leading to 
\be
\label{9.4}
\bt_k(x) = \sum_{n=0}^k b_n x^n \; .
\ee
 
Applying the method of self-similar approximants to expansion (9.4), as has 
been done above, we get a self-similar approximant
$\beta_k^*(x)$ whose limit, being by definition finite,
$$
 \bt_k^*(x) \ra const \qquad ( x \ra \infty ) \;  ,
$$
gives us the sought approximate expression for the critical exponent 
\be
\label{9.5}
 \bt_k^* = \lim_{x\ra\infty} \bt_k^*(x) \;  .
\ee

Note that the value of the critical amplitude $B$ does not need to be
considered at all. Below, we illustrate this method of calculating the
critical exponents by concrete examples.

\subsection{One-dimensional anharmonic oscillator}

Let us consider, as in Sec. 6.2, the model of the one-dimensional anharmonic 
oscillator whose mathematical structure is typical for many applied 
problems, yielding strongly divergent perturbation series.

The exact critical exponent, as follows from Eq. (6.7), is
$$
 \bt = \frac{1}{3} \;  .
$$
In addition to the coefficients $c_n$ of Sec. 6.2, we shall analyze the 
higher-order terms of sum (6.6), with the coefficients
$$
c_5 = \frac{916731}{256} \; , \qquad 
c_6 = - \; \frac{65518401}{1024} \; , \qquad
c_7 = \frac{2723294673}{2048} \; ,
$$
$$
 c_8 = -\; \frac{1030495099053}{32786} \; , \qquad
c_9 = \frac{54626982511455}{65536} \; , \qquad
c_{10} = -24478940702.8  \; .
$$

Employing the scheme of Sec. 9.1, we find, for the critical exponent, the 
factor approximants $\beta_4^*=0.241$, $\beta_7^*=0.303$, and $\beta_8^*=0.282$. 
Iterated roots result in $\beta_2^*=0.397$, $\beta_3^*=0.181$, but $\beta_4^*$
is complex. Corrected iterated roots yield $\beta_{2/2}^* = 0.307$, 
$\beta_{2/3}^* = 0.328$, $\beta_{2/4}^* = 0.310$, $\beta_{2/5}^* = 0.346$, and 
$\beta_{2/6}^* = 0.305$. Power-transformed roots give two solutions, $0.156$ 
and $0.238$. Doubly renormalized iterated roots of second order lead to 
$0.319$. As we see, the self-similar approximants are rather accurate, being 
close to $0.3$.

\subsection{Three-dimensional polymer coil}

As another example, we consider the three-dimensional polymer coil of Sec. 8.5.
The exponent found numerically, according to Eq. (8.14), is 
$$
 \bt=0.3544 \;  .
$$

Following the scheme of Sec. 9.1, we obtain the self-similar approximants 
for the critical exponent. Factor approximants yield $\beta_3^* = 0.343$, 
$\beta_4^* = 0.346$, and $\beta_5^* = 0.349$. Iterated roots result in 
$\beta_2^* = 0.345$, $\beta_3^* = 0.343$, $\beta_4^* = 0.351$, and 
$\beta_5^* = 0.349$. Power-transformed iterated roots give two solutions, 
$0.285$ and $0.349$ and corrected iterated roots give $\beta_{1/4}^* = 0.348$, 
$\beta_{2/2}^* = 0.345$, $\beta_{3/2}^* = 0.349$. Doubly-renormalized 
iterated roots yield $\beta_4^{**} = 0.353$, $\beta_5^{**} = 0.355$. All 
these approximants are close to the numerical value $\beta = 0.3544$.

\section{Equation of state}

The problems, considered in previous sections, were related to the cases
when it was necessary to find the large-variable behaviour of the studied 
functions. However, generally, self-similar approximation theory allows us to
derive approximants valid for the whole range of the variable. To illustrate 
this, we show below how it is possible to construct an equation of state, 
providing a good description in the whole region of densities.

Let us consider a system of quantum hard spheres \cite{Kel1996} characterized
by the $s$-wave scattering length $a_s$ corresponding to the diameter of a
hard sphere. The ground-state energy, in the limit of low density $\rho \ra 0$, 
is given \cite{Lee1957} by the asymptotic expression
\be
\label{10.1} 
 \frac{E}{N} \simeq 2\pi\; \frac{\rho a_s}{m} \left ( 1 +
\frac{128}{15\sqrt{\pi}} \; \sqrt{\rho a_s^3 } \right ) \;  ,
\ee
where $m$ is a sphere mass. The density can increase up to the value $\rho_0$,
when the system of the spheres becomes close packed. For a primitive hexagonal
close packing, such as producing a face-centred cubic arrangement, 
\be
\label{10.2}
  \rho_0 = \frac{\sqrt{2}}{a_s^3} \; .
\ee
In the close-packed limit, the energy behaves as
\be
\label{10.3}
  \frac{E}{N} \simeq \frac{B}{2m} \left ( \rho^{-1/3} - \rho_0^{-1/3}
\right ) \; ,
\ee
with the experimental value
\be
\label{10.4}
B \equiv 2^{2/3} \pi^2
\ee
found by Cole \cite{Col1967}.

To rewrite the low-density asymptotic expression in a more convenient way, we
introduce the variable $x$ by the relation
\be
\label{10.5}
 \frac{\rho}{\rho_0} = \frac{x^6}{(1+x^2)^3} \;  .
\ee
As is seen, $x \ra 0$ when $\rho \ra 0$ and $x \ra \infty$ when $\rho \ra \rho_0$. 
With the new variable, expansion (10.1) for $x \ra 0$ takes the form
\be
\label{10.6}
 \frac{E}{N} \simeq  2\pi\; \frac{\rho_0 a_s}{m} \; x^6 \left ( 1 -
3x^2 + \frac{128}{15\sqrt{\pi}} \; \sqrt{\rho_0 a_s^3 }\; x^3 + 6x^4 - \;
\frac{192}{5} \; \sqrt{\rho_0 a_s^3} \; x^5 \right ) \; ,
\ee
while the close-packed limit reads as
\be
\label{10.7}
  \frac{E}{N} \simeq  \frac{\pi^2}{ma_s^2} \; x^4 \qquad (x\ra\infty) \; .
\ee
Using the iterated root approximant of second order for expansion (10.6), we get
\be
\label{10.8}
 \frac{E^*_2}{N} = 2\pi\; \frac{\rho_0 a_s}{m} \; x^6 \left ( 
1 + A_2 x^2 \right )^{-1} \;   ,
\ee
with $A_2 = 2 \sqrt{2} / \pi$ corresponding to limit (10.7). Inverting the change
of the variable (10.5), we return to the initial variable, that is, to density,
obtaining the equation of state
\be
\label{10.9}
 \frac{E^*_2}{N} = 2\pi\; \frac{\rho a_s}{m} \left [ 1 - \left (
\frac{\rho}{\rho_0} \right )^{1/3} \right ]^{-2}
\left [ 1 + b \left (\frac{\rho}{\rho_0} \right )^{1/3} \right ]^{-1} \; ,
\ee
in which
\be
\label{10.10}
b = \frac{2\sqrt{2}}{\pi} - 1 \;   .
\ee
This equation {\it exactly} coincides with the empirical equation called the 
modified London equation \cite{Sol1994} that is in very good agreement with the
Green function Monte Carlo computer simulations for the many-body hard-sphere
fluid \cite{Kal1974}. Higher orders of the self-similar iterated root approximants,
as we have checked, do not essentially change the accuracy of the equation of 
state (10.9) that already gives a perfect agreement with computer simulations.

\section{Conclusion}

We have considered the problem of extrapolating perturbation-theory expansions,
obtained for asymptotically small variable $x \ra 0$, to the large-variable
limit $x \ra \infty$. For this purpose, we have applied the theory of 
self-similar approximations, concentrating on six different variants, resulting 
in self-similar factor approximants (Sec. 3.1), self-similar root approximants 
(Sec. 3.2), iterated root approximants (Sec. 3.3), corrected root approximants
(Sec. 3.4), self-similar power-transformed approximants (Sec. 3.5), and doubly
renormalized self-similar approximants (Sec. 3.6).

Pad\'{e} approximants are shown to be much less accurate than the self-similar 
approximants, and often not applicable at all. In some cases, more refined 
techniques, such as the Kleinert variational perturbation theory, employing 
control functions introduced through a variable transformation, can give better 
accuracy, although they are essentially more complicated. However, our main aim 
here has been the analysis of the validity of the approximants that could provide 
good accuracy, at the same time being sufficiently simple for calculations and 
yielding explicit analytical formulas. 
 
In order to demonstrate the wide applicability of the self-similar approximants,
we treated a number of examples of rather different nature. In the majority 
of cases, the approximants yield close results and provide good accuracy of 
extrapolation. In general, their accuracy is essentially higher than that of 
Pad\'{e} approximants. In some cases, the latter are not applicable at all, 
giving qualitatively wrong results, while self-similar approximants do work in
such cases. 

Comparing different variants of the analyzed self-similar approximants, we see
that power-transformed approximants often lead to multiple solutions for the 
sought parameters, because of which they are less convenient than other 
approximants enjoying unique solutions. The doubly renormalized approximants, 
although improving the final results, are cumbersome allowing only for their 
complicated numerical calculation. The self-similar factor approximants and 
iterated root approximants seem to be the most convenient for the purpose of
the considered extrapolation. 

Having to hand several methods of self-similar extrapolation is important
because of the following reason. A problem under consideration can be so 
complicated that the exact answer is not known and only a few terms of 
perturbation theory are available, then it is rather difficult to judge the 
accuracy of the approximation used. However, if different methods give close 
results, this serves as an argument that the obtained approximations are 
reliable. 

Finally, we have considered problems whose large-variable behaviour is
of power-law type. We are aware that there exists another class of problems
possessing exponential behaviour and also demonstrating the Stokes phenomenon.
For the problems of this class, it is necessary to use another variant of the
self-similar approximation theory, involving self-similar exponential
approximants \cite{YukGlu1998, Yuk1998}. These, as has been demonstrated 
in the cited papers, make it possible to derive accurate approximations for
the functions of exponential behaviour as well as to treat problems 
accompanied by the Stokes phenomenon. We do not address such problems here 
but they have been studied in our previous publications \cite{YukGlu1998, Yuk1998}.

\vskip 1cm
{\bf Acknowledgement}

\vskip 3mm

One of the authors (V.I.Y.) is grateful for useful discussions to E.P. 
Yukalova and for financial support, to the Russian Foundation for Basic 
Research.

\end{document}